\documentclass[pdftex,twocolumn,epjc3]{svjour3}
\RequirePackage[T1]{fontenc}
\smartqed  

\RequirePackage{graphicx}
\RequirePackage{mathptmx}      
\RequirePackage{flushend}
\RequirePackage[numbers,sort&compress]{natbib}
\RequirePackage[colorlinks,citecolor=blue,urlcolor=blue,linkcolor=blue]{hyperref}

\journalname{Eur. Phys. J. C}
 
\usepackage{graphicx,slashed,bbold,subfigure,amsmath,amssymb,hyperref}
\usepackage[utf8]{inputenc}
\usepackage[english]{babel}

\def\ca{\c{c}\~{a}}

\begin{document}
\title{Top condensation model: a step towards the correct prediction of the Higgs mass}

\author{A. A. Osipov\thanksref{e1,addr1}
\and B. Hiller\thanksref{e2,addr2}
\and A. H. Blin\thanksref{e3,addr2}
\and F. Palanca\thanksref{e4,addr2}
\and J. Moreira\thanksref{e5,addr2}
\and M. Sampaio\thanksref{e6,addr5}}

\thankstext{e1}{e-mail: aaosipov@jinr.ru}
\thankstext{e2}{e-mail: brigitte@fis.uc.pt}
\thankstext{e3}{e-mail: alex@uc.pt}
\thankstext{e4}{e-mail: fpalanca2@gmail.com}
\thankstext{e5}{e-mail: jmoreira@uc.pt}
\thankstext{e6}{e-mail: msampaio2@yahoo.com}

\institute{Bogoliubov Laboratory of Theoretical Physics, JINR, Dubna, Moscow Region,Russia\label{addr1}
\and CFisUC, Department of Physics, University of Coimbra, P-3004-516 Coimbra, Portugal\label{addr2}
\and ABC Federal U., Santo André, São Paulo, Brazil\label{addr5}}

\date{Received: date / Accepted: date}
\maketitle

\begin{abstract}
A realization of the composite Higgs scenario in the context of the effective model with the $SU(2)_L\times U(1)_R$ symmetric four-Fermi interactions proposed by Miransky, Tanabashi and Yamawaki is studied. The model implements  Nambu's mechanism of dynamical electroweak symmetry breaking leading to the formation of  $\bar tt$ and $\bar bb$ quark condensates. We explore the vacuum structure and spectrum of the model by using the Schwinger proper-time method. As a direct consequence of this mechanism, the Higgs acquires a mass in accord with its experimental value. The present prediction essentially differs from the known overestimated value, $m_\chi= 2m_t$, making more favourable the top condensation scenario presented here. The mass formulas for the members of the second Higgs doublet are also obtained. The Nambu sum rule is discussed. It is shown that the anomalous $U(1)_A$ symmetry breaking modifies this rule at next to leading order in $1/N_c$.  
\end{abstract}

\maketitle
\section{Introduction}
The top-condensation models have already a long history of development \cite{Terazawa1977,Terazawa1980a,Terazawa1980b,Nambu88,Nambu89,Mir89,Yam89,Bardeen90,Luty90,Suzuki90,Harada91,Cvetic99,Hill03,Yam16}. The main idea of this approach is that the Higgs sector of the Standard Model (SM) is originated by the four-Fermi quark couplings which describe physics below a physical cutoff $\Lambda$. These models are an attractive framework to study the origin of mass beyond the Standard Model (SM), but reportedly suffer from the following phenomenological problem: the predicted mass of the standard Higgs $\chi$ is too large $m_\chi =2m_t$, where $m_t=173\,\mbox{GeV}$ is the top quark mass. This result, obtained in the large $N_c$ approximation \cite{Bardeen90} in the one Higgs doublet model, survives in the two-Higgs-doublet models studied in \cite{Luty90,Suzuki90}. The inclusion of the effects due to the gauge and Higgs contributions \cite{Bardeen90,Luty90} computed through the one-loop renormalization-group equations does not essentially improve the situation. The Higgs mass obtained is still large $m_\chi \simeq \sqrt 2 m_t$. A number of attempts to resolve the Higgs mass problem at large $N_c$ by using the top-seesaw mechanism have been made in \cite{Fukano12,Fukano14,Fukano15}. In these approaches the light Higgs emerges as one of the composite pseudo Nambu-Goldstone bosons associated with the spontaneous breaking of an approximate global $U(3)_L\times U(1)_R$ symmetry down to $U(2)_L\times U(1)_V$ by the corresponding quark condensates. 

The result $m_\chi =2m_t$ is an analogue of the relation found in the Nambu-Jona-Lasinio (NJL) model \cite{NJL1,NJL2} for the mass of the scalar fermion-antifermion bound state (so called $\sigma$-meson) $m_\sigma = 2m_f$, where $m_f$ is the mass of the fermion in the Nambu-Goldstone phase. The generalized form of this relation is known as the Nambu sum rule \cite{Nambu,Volovik13,Volovik}. According to this hypothesis, collective bosonic modes arising in a system with four-Fermi interactions can be combined into pairs (the so-called Nambu partners) for each of which the equality $m_1^2 + m_2^2 = 4m_f^2$ is fulfilled. This formula relates the gap in the fermion spectrum, $m_f$, with the corresponding gaps in the boson spectrum $m_1$ and $m_2$. It follows from the Nambu sum rule that a phenomenologically acceptable result $m_\chi\simeq m_t /\sqrt 2$ can be obtained only in the extended version of the top-condensation model containing at least two doublets of Higgs fields. In the latter case the sum rule suggests the existence of a Nambu partner for the standard Higgs boson with a mass around $325\,\mbox{GeV}$ \cite{Volovik}. It is generally believed (see e.g., \cite{Soni14,Yam15}) that the  conventional top-condensation models with two Higgs doublets \cite{Mir89,Luty90,Suzuki90} cannot reproduce this result, because the mass $m_\chi$ tends to be in the range $m_t < m_{\chi}\leq 2m_t$. 

In this paper, contrary to above mentioned belief, we show that in the model \cite{Mir89} the mass $m_\chi$ varies in the range $2m_b < m_{\chi}< m_t$, and as a consequence a light SM-like Higgs with a mass of $\mathcal{O}(m_W)$ emerges. Our conclusion is based on the spectrum, which follows from the Schwinger-DeWitt expansion \cite{Schwinger51,DeWitt65,DeWitt75,Ball89} of the effective action of the model \cite{Mir89} at low-energies $\mu \sim \Lambda_{EW}=(\sqrt 2 G_F)^{-1/2}=246\,\mbox{GeV} \ll\Lambda .$ 

Note that the four-quark terms at the scale $\Lambda $ can be written in terms of static boson fields with Yukawa couplings to the quarks. This form reproduces the four-Fermi couplings when auxiliary fields are integrated out. On the other hand, in the large-$N_c$ limit, the theory  evolves at low energy to the extended SM with the dynamically generated two Higgs doublet fields. It accounts for the one-quark-loop contribution where one keeps only the leading quadratic and logarithmic divergences in the form suggested in \cite{Bardeen90}. It is interesting that the obtained composite Higgs theory can be identified with particular renormalization-group trajectories known from the two Higgs-doublets models \cite{Branco12}.

The implementation of the Nambu sum rule for the model \cite{Mir89} is also considered. We show that the anomalous breaking of the $U(1)_A$ symmetry modifies this rule. The two Higgs doublets of the model contain eight real fields, three of which are absorbed by the gauge $W^\pm$ and $Z$ bosons due to the Higgs mechanism. Of the remaining five fields, two charged states, $\chi^\pm$, are Nambu partners. However, the other three neutral states $\chi_0, \chi_3$, and $\phi_0$ cannot be reduced to the picture with two Nambu partners, and as a result, the sum rule takes a slightly different form, not directly relating the masses of Higgs states to the gap in the fermion spectrum. We argue that this modification is the result of going beyond the leading order approximation in $1/N_c$ already in the four-Fermi Lagrangian\footnote{The large $N_c$ behaviour of the model is reflected in the coupling constants, in particular, it is well established that the NJL four-quark interaction dominates over the 't Hooft $U(1)_A$ breaking interaction as $N_c^{N_f-1}$, where $N_f$ is the number of flavors (in our case $N_f=2$).}, where the 't Hooft interaction violating the Nambu sum rule is $1/N_c$ suppressed.

The paper is organized as follows. In Sec.II we present the most important features of the model \cite{Mir89}. The new aspect of our approach here is the derivation of the low-energy effective action of the model on the basis of the Schwinger-DeWitt background field method. Through the asymptotic proper time expansion we obtain the gap equations, spectrum and coupling constants of the theory. We also obtain and discuss the Nambu sum rule. Here we give the exact form of these relations which one obtains in the model and show the role of the $U(1)_A$ anomaly in the modification of this sum rule. The numerical estimations are given in Sec.III. The discussion of the hierarchy problem is given in the end of this section. We summarize our results in Sec. IV. The six appendices contain important details of our calculations.      

\section{The top-bottom system with four-quark interactions}
\label{4qLInt}

We discuss below the general features of the model \cite{Mir89} where for simplicity only the quarks of the third generation are considered. This includes the dynamical symmetry breaking through the top-quark condensation, the spectrum of the Higgs and gauge-boson states, the implementation of the Nambu sum rule. Our conclusions here are based on the low-energy effective Lagrangian obtained by the Schwinger -- DeWitt expansion, and the gap equations, which describe the ground state of the theory. We adhere to the semiclassical approximation, i.e., the quantum corrections due to gauge, quark or composite Higgs fields are not considered.

\subsection{Effective Lagrangian}

Let us consider the quark-gauge-boson system described at very large scale $\Lambda$ (perhaps of the order of the grand-unified-theory (GUT) scale $10^{15}$ GeV) by the $SU(2)_L\times U(1)_R$ gauge symmetric Lagrangian density
\begin{align}
\label{stL1}
\mathcal L&=\bar \psi_L i\gamma^\mu \mathcal D_\mu \psi_L + \sum_{a=1}^2\bar \psi^a_{R} i\gamma^\mu \mathcal D_\mu \psi^a_{R} + \mathcal L_{\mbox{\tiny YM}} + \mathcal L_{4\psi}.
\end{align} 
Here we focus, for simplicity, upon the third electroweak generation of quarks. The color indices of quark fields are suppressed. The heavy quark $SU(2)$ doublet is
\begin{align}
\psi =\left( \begin{array}{c} 
\psi_1\\ \psi_2 \end{array} \right)=
\left( \begin{array}{c} t \\ b \end{array}\right).
\end{align}
The chiral right-left projection operators are defined as follows $P_R=\frac{1}{2}(1+\gamma_5)$, $P_L=\frac{1}{2}(1-\gamma_5)$ with $\psi_{L.R}=P_{L,R} \psi$. The gauge covariant derivatives have a standard form  
\begin{align}
\mathcal D_\mu \psi_L  &=\left(\partial_\mu -ig_RT_i A_{R\mu}^i -ig'_RY_LB_{R\mu} \right) \psi_L, \\
\mathcal D_\mu \psi_R  &=\left(\partial_\mu -ig'_RQB_{R\mu}\right)\psi_R,
\end{align}   
where the matrix $Q = T_3 + Y_L $ describes the electromagnetic charges of top and bottom quarks in relative units of the proton charge $e > 0$, $Y_L=1/6$; $A_{R\mu} = A_{R\mu}^iT_i$ and $B_{R\mu}$ are gauge fields\footnote{Here the symbol $R$ anticipates that in the following we may exercise the freedom of rescaling the gauge fields.} of the $SU(2)_L$ and $U(1)_R$ groups of local transformations, respectively; the $SU(2)$ Lie algebra generators are $T_i =\tau_i/2,\ i=1,2,3$, where $\tau_i$ are Pauli matrices. 

At this point it is useful to anticipate some of the discussion of Section \ref{Gauge} concerning the impact of induced quantum corrections due to the integration of the short-distance components of quark fields. This integration leads to the rescaling of the gauge fields and their couplings with quarks at low-energies. In particular, $g_R=Z_A^{1/2}g$ and $g'_R=Z_B^{1/2}g'$, where $g$ and $g'$ are renormalized induced coupling constants, the parameters $Z_A$ and $Z_B$ are renormalization factors, which are needed at the low-energy scale $\mu\ll \Lambda$ to have an entire correspondence with the Yang-Mills part of the SM. Notice that $g'/g=\tan\theta_W$, $e=g\sin\theta_W$, $\sin^2\theta_W=0.23$. As it will be shown later, $Z_A$ and $Z_B$ depend upon $\Lambda$, and $\mu$; they are equal to one at $\mu = \Lambda$.

The pure Yang-Mills part of the Lagrangian density $\mathcal L_{\mbox{\tiny YM}}$ is standard
\begin{align}
\label{YM}
\mathcal L_{\mbox{\tiny YM}}=-\frac{1}{4}\left(B_{R\mu\nu}^2 + \vec G_{R\mu\nu}^2 \right),
\end{align}   
where
\begin{align}
B_{R\mu\nu}&=\partial_\mu B_{R\nu} -\partial_\nu B_{R\mu}  , \\
\vec G_{R\mu\nu}&=\partial_\mu \vec A_{R\nu} -\partial_\nu \vec A_{R\mu} +g_R \vec A_{R\mu} \times \vec A_{R\nu}.
\end{align}
The physical gauge fields $\vec A_\mu$, $B_\mu$ are defined through the rescaling $\vec A_{R\mu}=Z_A^{-1/2}\vec A_\mu$, $B_{R\mu}=Z_B^{-1/2} B_\mu$.    

The Lagrangian density (\ref{stL1}) does not have the electroweak Higgs sector. Instead, it is assumed that at some high energy scale $\Lambda$ the local four-quark interaction $\mathcal {L}_{4\psi}$ arises as an effective description for new physics beyond the SM. This local interaction has the most general $SU(2)_L\times U(1)_R$ invariant form \cite{Mir89}
\begin{align}
\label{MTY}
\mathcal {L}_{4\psi} =&
g_1\left(\bar \psi_L^a\psi_R^b\right)\left(\bar \psi_R^b\psi_L^a\right)\nonumber \\
+&g_2\left(\bar \psi_L^a\psi_R^b\right)\left(i\tau_2\right)^{ac}\left(i\tau_2\right)^{be}\left(\bar \psi_L^c\psi_R^e\right)\nonumber\\
+&g_3\left(\bar \psi_L^a\psi_R^b\right)\tau_3^{bc}\left(\bar \psi_R^c\psi_L^a\right)+\mathrm{h.c.},
\end{align}
where the three independent couplings $g_i$ are assumed to be real and have the same dimension $[g_i]=M^{-2}$. These four-fermion operators are the lowest mass dimension operators we can add to the SM using only quark fields of the third generation. We discuss the model with real and positive $g_i$ hereafter. As in equation (\ref{stL1}), the color degrees of freedom of quarks are not explicitly indicated in (\ref{MTY}), but it is assumed that implicit summation is carried out between the fields enclosed in parentheses. 

The individual terms of the four-quark Lagrangian possess the following symmetry
\begin{align}
g_1: &\quad SU(3)_c\times SU(2)_L\times SU(2)_R\times U(1)_V\times
  U(1)_A. \nonumber \\
g_2: &\quad SU(3)_c\times SU(2)_L\times SU(2)_R\times U(1)_V. \\
g_3: &\quad SU(3)_c\times SU(2)_L\times U(1)_R\times U(1)_V\times
  U(1)_A. \nonumber 
\end{align}
The theory with only one non-zero coupling $g_1$ would have a spectrum typical for the NJL-model with $U(2)_L\times U(2)_R$ chiral symmetry, which leads in the Nambu-Goldstone phase to structured bosons, four massless pseudoscalars and four mass-degenerate scalar states with a mass twice the induced fermion mass. In the absence of the term $\propto g_2$, the Lagrangian density (\ref{stL1}) has a global $U(1)_A$ symmetry, which being broken spontaneously leads to a massless Nambu-Goldstone mode. Such a boson is not observed experimentally, therefore, $g_2\neq 0$. The interaction with the coupling constant $g_3$ violates spatial parity and isotopic symmetry. 

\subsection{Bosonic variables }

The description of collective bound states can be facilitated if we introduce in the Lagrangian density (\ref{stL1}) eight auxiliary boson fields $\sigma = \sigma_\alpha \tau_\alpha ,\, \pi =\pi_\alpha \tau_\alpha$\footnote{It is assumed here and hereinafter that the Greek index $\alpha $ runs through the values $ \alpha = 0,1,2,3 $, and $ \tau_0 = 1 $.}. With these static variables the fermion part of (\ref{stL1}) takes the Yukawa form (see  \ref{app1} for details) 
\begin{equation}
\label{sembos}
\mathcal L' = \bar\psi \left(i\gamma^\mu \mathcal D_\mu +\sigma +i\gamma_5 \pi
  \right) \psi +  \mathcal {L}_{\pi ,\sigma} +\mathcal L_{\mbox{\tiny YM}},
\end{equation}
where $\mathcal {L}_{\pi ,\sigma}$ is quadratic in the boson fields
\begin{align}
\label{b4q}
\mathcal {L}_{\pi ,\sigma} =&-\frac{1}{\bar g^2}\left[(g_1+g_2)
  (\pi_0^2+\sigma_i^2) +(g_1-g_2) (\sigma_0^2+\pi_i^2) \right. \nonumber \\
&\left. -2g_3(\pi_0\pi_3 +\sigma_0\sigma_3 -\sigma_1\pi_2+\sigma_2\pi_1)  \right], 
\end{align}
with $\bar g^2\equiv g_1^2-g_2^2-g_3^2\neq 0$, and the electroweak gauge covariant derivative is given by
\begin{align}
\mathcal D_\mu\psi = \left[\partial_\mu -igT_iA_\mu^i P_L +ig' B_\mu \left( T_3P_L -Q \right) \right]\psi .
\end{align}
 
We wish to emphasize that this form of the Lagrangian density has the same dynamical content as
(\ref{stL1}). This point is clarified by solving the Euler-Lagrange equations of static fields following from (\ref{sembos}). In \ref{app1} we present details of such calculations. 

The Lagrangian density (\ref{sembos}) contains the same number of parameters as (\ref{stL1}). We refer to this scenario as minimal bosonization. As it has been pointed out by T. Eguchi \cite{Eguchi:78}, the four-fermion theories can induce new coupling constants, which become independent of the original four-fermion couplings. This means that in principle one can write another Lagrangian density, $\mathcal L''$, which will contain additional new parameters and still will be dynamically equivalent to (\ref{stL1}). Our goals here can be achieved already with the Lagrangian density (\ref{sembos}), to which we restrict ourselves in this work, so as not to complicate the calculations.

The static fields $\sigma_\alpha$ and $\pi_\alpha$ acquire kinetic terms and self-interactions, provided that Yukawa couplings, and the low-energy scale $\mu$ are tuned so that the couplings of four-quark interactions $g_1, g_2, g_3$ are near their critical value. Let us consider this point in detail.

\subsection{Schwinger-DeWitt expansion}

The model we are considering can be described by a generating functional
\begin{align}
\label{gf}
Z =\!\!\int\!\! d\sigma_\alpha d\pi_\alpha d\psi d\bar\psi \exp \left(
    i\!\!\int\!\! d^4x \mathcal L' \right). 
\end{align}
As we noted  above, the integration over boson fields in (\ref{gf}) will return us to the initial theory with the Lagrangian density (\ref{stL1}). On the other hand, the path integral representation is convenient for the $1/N_c$ expansion of the theory, since here it is already possible to integrate out the short-distance components of the quark fields $\psi\to\psi +\psi_{sd}$ by using Wilson's method \cite{Wilson74}. Indeed, the analyses of \cite{Bardeen90} may be interpreted as implying that at scales below the cutoff $\Lambda$ the auxiliary scalar fields $\sigma$ and $\pi$ develop induced, fully gauge invariant kinetic terms and quartic interaction contributions in the effective action. Here we derive these induced terms in the case of the model with Lagrangian density (\ref{sembos}) by using the Schwinger-DeWitt technique \cite{Schwinger51,DeWitt65,DeWitt75,Ball89} with the specially assigned cutoff procedure. Notice that the Lagrangian density (\ref{b4q}) requires a diagonalization. However, we calculate the induced effective Lagrangian part first, since it also contributes to the non-diagonal quadratic form. 

The full induced contribution of the integrated short-distance quark components to the real part of the effective action can be represented in the form of the asymptotic series in proper time $t$ (in the Euclidean space, as indicated by the symbol "$E$")
\begin{align}
\label{SD}
\mbox{Re}\, S_E=-\frac{1}{2}\int\! \frac{dt}{t^3} \int\! d^4x_E
  \frac{1}{(4\pi )^2} \sum_{n=0}^\infty t^n \mbox{tr} (a^E_n),
\end{align}
where $a_n^E$ are the Seeley-DeWitt coefficients that depend on fields and, in particular, are $a^E_0=1$, $a^E_1=-Y$, $a^E_2=Y^2/2-F_{\mu\nu}^{2}/12$, with $Y$ defined below in eq. (\ref{Y}). The remaining coefficients are not required for our purpose, because only for $n = 0, 1, 2$ do the integrals in proper time diverge and therefore make the dominant contribution at low energies. Since $a_0$ does not contain fields, we need to consider only two integrals, which we denote as $C_1$ and $C_2$ and define by introducing two scale parameters $\Lambda$ and $\mu$, 
\begin{align}
\label{C}
C_1 & = \int_{1/\Lambda^2}^{1/\mu^2} \!\frac{dt}{t^2} =
      \Lambda^2-\mu^2, \\
C_2 & = \int_{1/\Lambda^2}^{1/\mu^2} \!\frac{dt}{t} = \ln \frac{\Lambda^2}{\mu^2}.
\end{align}
The parameter $\Lambda$, as we have already noted, makes sense of the scale at which unknown physics is approximated by effective four-quark interactions (\ref{MTY}). The second parameter $\mu$ is a low-energy scale, relative to which one-loop contributions are determined. This is done in such a way that at the large scale $\mu =\Lambda$ all contributions induced by the proper time expansion become zero, as it is required \cite{Bardeen90}.

In the model considered, one obtains the following expressions for the field dependent functions in $a_n^E$
\begin{align}
\label{Y}
&Y  = \sigma^2+\pi^2 +i\gamma_5 [\sigma, \pi ] -i\nabla_\mu 
  (\gamma^\mu\sigma +i\gamma^\mu\gamma_5\pi )\nonumber \\
& \ \ \   -\frac{i}{4}  [\gamma_\mu , \gamma_\nu ] F^{\mu\nu}, \\
\label{F}
&F_{\mu\nu}=\partial_\mu \Gamma_\nu-\partial_\nu \Gamma_\mu
  -i[\Gamma_\mu ,  \Gamma_\nu ], \\
\label{Gm}  
&\Gamma_\mu =gT_i A_\mu^i P_L-g'B_\mu (P_LT_3-Q).
\end{align}
These functions are given already after its extrapolation to the Minkowski space, the necessary ingredients for its derivation can be found, for instance, in \cite{Osipov:17}. 

As a consequence of the fact that $\Gamma_\mu$ depends on the $\gamma_5$ matrix, one should be careful applying the covariant derivatives to the scalar fields 
\begin{align}
\label{deriv}
&\nabla_\mu (\gamma^\mu\sigma ) = \partial_\mu (\gamma^\mu\sigma ) -i[\Gamma_\mu ,  \gamma^\mu \sigma ],  \\
&\nabla_\mu (i\gamma^\mu\gamma_5\pi )=\partial_\mu (i\gamma^\mu\gamma_5 \pi ) -i[\Gamma_\mu ,  i\gamma^\mu \gamma_5 \pi ]. 
\end{align}
Thus, taking into account the leading divergencies in the proper time expansion of one-quark-loop contributions at low energies $\mu \ll \Lambda$ there appears an additional term described by the Lagrangian density 
\begin{align}
\label{dl}
\Delta \mathcal L_{sd} =  -\frac{1}{32\pi^2} \left[ C_1 \mbox{tr} (-Y) +C_2
  \mbox{tr}\left(\frac{Y^2}{2} -\frac{F^2_{\mu\nu}}{12} \right) \right],
\end{align}
where traces are calculated from the products of Pauli matrices, and over color and Dirac indices.  
 
Hence, the low-energy theory of fermions and bosons is described by the $ SU(2)_L \times U(1)_R$ gauge-invariant Lagrangian density
\begin{align}
\label{stL2}
\tilde {\mathcal L}= \mathcal L' +\Delta \mathcal L_{sd} .
\end{align} 
The last term does not change the original theory at high energies, because $\Delta\mathcal L_{sd} = 0 $ for $\mu = \Lambda$. However it becomes important at low energies $\mu\ll \Lambda$, inducing the potential of the composite Higgs particles, their interactions with gauge fields and kinetic terms of free bosonic fields.

\subsection{Higgs sector and the gap equation}

Let us consider the Higgs sector of the model. First we discuss the vacuum structure of the scalar potential, and derive scalar masses. From eq. (\ref{stL2}) it follows that the Higgs potential has the form
\begin{align}
\label{VH}
&V_H =-\mathcal L_{\pi, \sigma} -\bar C_1\left(\sigma_\alpha^2+\pi_\alpha^2 \right) + 2\bar C_2\left[\frac{1}{4}\left(\sigma_\alpha^2+\pi_\alpha^2 \right)^2 \right. \nonumber \\
&\left. +\left(\sigma_0^2+\pi_i^2 \right) \left(\sigma_i^2+\pi_0^2 \right) - \left(\sigma_0\pi_0 -\sigma_i \pi_i\right)^2 \right].
\end{align}
where $\bar C_{1,2}\equiv N_c C_{1,2}/(4\pi^2)$. 

This formula can be written in a more compact form if one introduces two electroweak doublets with the $U(1)$ hypercharge $Y_L=1/2$  (see \ref{app2})
\begin{align}
\label{phidoublets}
\Phi_1= {\pi_2+i\pi_1 \choose \sigma_0-i\pi_3}, \quad
\Phi_2= {\sigma_1-i\sigma_2 \choose -\sigma_3+i\pi_0}.
\end{align}  
These states are not physical because they do not correspond to the mass eigenstates. Nonetheless, it is useful to compare eq. (\ref{VH}), written in terms of these doublets, with the most general CP conserved scalar potential, the quartic part of which is symmetric under separate discrete transformations $\Phi_1\to -\Phi_1$, or $\Phi_2\to -\Phi_2$ given in \cite{Branco12} (see eq.(2) of that paper). In our case, we have  
\begin{align}
\label{H}
&V_H =- \bar C_1(\Phi_1^\dagger\Phi_1 +\Phi_2^\dagger\Phi_2) +2\bar C_2         \\
&\left[ \frac{1}{4}(\Phi_1^\dagger\Phi_1+\Phi_2^\dagger\Phi_2)^2 +(\Phi_1^\dagger
    \Phi_1)(\Phi_2^\dagger\Phi_2) - (\mbox{Im}(\Phi_1^\dagger\Phi_2) )^2\right] \nonumber  \\
& +\frac{1}{\bar g^2}\left[(g_1\!-\!g_2)\Phi_1^\dagger\Phi_1  +(g_1\!+\!g_2) \Phi_2^\dagger\Phi_2 +2g_3 \mbox{Re}(\Phi_1^\dagger\Phi_2)  \right]. \nonumber 
\end{align}
The comparison yields in the notation of \cite{Branco12}
\begin{align}
&m_{11}^2=\frac{g_1-g_2}{\bar g^2} - \bar C_1, \quad m_{22}^2=\frac{g_1+g_2}{\bar g^2} - \bar C_1, \quad m_{12}^2=-\frac{g_3}{\bar g^2}, \nonumber \\
&\lambda_1=\lambda_2=\frac{1}{3}\lambda_3=-\lambda_4=\lambda_5=\bar C_2.  
\end{align}
It shows that the potential of the model considered is quite restrictive, it has only four real independent parameters instead of eight in the most general case for such class of models. This is a direct consequence of the approximation made in our step from the Lagrangian density (\ref{stL1}) to the model of "minimal bosonization" with the Lagrangian density (\ref{sembos}).

Assuming that the vacuum expectation values of the fields $\sigma_0$ and $\sigma_3$ differ from zero: $\langle \sigma_0 \rangle\! = \! -m_0 $, $ \langle \sigma_3 \rangle \! = \! -m_3$, one finds the minimum conditions for the potential energy (gap equations) to determine $m_0$ and $m_3$
\begin{align}
\label{extr1}
&m_0 (g_1\!-\!g_2)-m_3 g_3=\bar g^2 m_0 [\bar C_1-(m_0^2+3m_3^2)\bar C_2 ],\\
\label{extr2}
&m_3 (g_1\!+\!g_2)-m_0 g_3=\bar g^2 m_3  [\bar C_1-(m_3^2+3m_0^2)\bar C_2 ].
 \end{align} 
The fulfilment of gap equations guarantees the absence in $V_H$ of linear in $\sigma_0$ and $\sigma_3$ terms at scale $\mu$ after the redefinition of these variables $\sigma_0\to\sigma_0-m_0$, $\sigma_3\to\sigma_3-m_3$. The nonzero vacuum expectation values lead to the gap in the fermion spectrum. As a result, the top and bottom quarks acquire the nonzero masses $m_t = m_0 + m_3$, $m_b = m_0-m_3$. A phenomenologically acceptable solution $m_t \gg m_b$ arises at $m_0 \simeq m_3 $. Details about the quark condensate content of these masses are given in \ref{app3}. 

Gap-equations rewritten in terms of quark masses have the following form
\begin{align}
\label{extrem1}
&(g_1\!-\!g_3)m_t - g_2m_b =\bar g^2 m_t (\bar C_1- m_t^2\bar C_2 ),\\
\label{extrem2}
&(g_1\!+\!g_3)m_b - g_2m_t=\bar g^2 m_b (\bar C_1- m_b^2 \bar C_2 ).
 \end{align}   
It is obvious that for $\mu=\Lambda$ this system has only a trivial solution $m_t=m_b=0$: the condition $\bar g^2\neq 0$ warrants the absence of a nontrivial solution. In the region $\mu <\Lambda$, the right-hand side of this system differs from zero and, in the strong coupling regime, the equations can possess nontrivial solutions. 

The easiest way to establish this fact is to consider a particular case $g_2=0$. In this case eqs. (\ref{extrem1})-(\ref{extrem2}) are decoupled with respect to the quark masses
\begin{align}
\label{extr1c}
&m_b^2\,\bar C_2 =\bar C_1 -  \frac{1}{g_1-g_3}\, ,   \\
\label{extr2c}
&m_t^2\,\bar C_2 =\bar C_1 -  \frac{1}{g_1+g_3}\, .
\end{align} 
It follows then that a bottom quark may acquire its mass even if $g_2 = 0$. Indeed, one easily comes to the inequality $g_1-g_3>g_c=4\pi^2/N_c\Lambda^2$ under which we can expect that eq. (\ref{extr1c}) has a nontrivial solution. This differs from the model \cite{Harada91}, where $g_2\neq 0$ is a main requirement for the bottom quark mass to take non zero value. The trivial solution, $m_b=0$, exists only if $g_2=0$ as it follows from eq. (\ref{extrem2}). Note that in the case $g_2=g_1=0$, the theory is reduced to the standard version of the NJL model. As is known, if the constant $g_1$ exceeds its critical value $g_c$, the vacuum of the model becomes unstable to the trivial solution $m_0=m_3=0$ and, as a consequence, the Fermi fields become massive.

Due to the hierarchy problem (as it will be shown below), the fulfilment of the inequality $g_1> g_c$ is a necessary condition for the generation of masses of top and bottom quarks. It will also be a sufficient condition, if the constant $g_3\neq 0$. This is because the $g_3$-introduced four-quark interaction is responsible for the difference of quark masses $m_t\neq m_b$. Therefore, we can speak of $\langle \bar bb\rangle$ condensate catalysis which takes place at any arbitrarily small value of the constant $g_3$. A similar statement was made in \cite{OsipovKh:19}, but in a slightly different context.

The quadratic part of $V_H $ can be diagonalized by two orthogonal transformations. One of them, characterized by the angle $\theta$, diagonalizes the charged sector. The other angle $\theta'$ is associated with the diagonalization of the neutral scalar states. The tangents of these angles are expressed in terms of the ratio $m_3/m_0$ and the ratio of the couplings $g_3/g_2$
\begin{align}
\label{ta}
\tan\theta =\frac{m_3}{m_0}, \quad \tan 2\theta'=3\tan 2\theta - 2\, \frac{g_3}{g_2}.
\end{align}
The rotations to the mass eigenstates are described by the following formulas (see \ref{app4})
\begin{align}
\Phi_1=  H_1\cos\theta + H_2 \sin\theta =\frac{1}{m}\left(m_0 H_1 +m_3 H_2 \right), \\
\Phi_2= H_2 \cos\theta  - H_1\sin\theta =\frac{1}{m}\left(m_0 H_2 -m_3 H_1 \right),
\end{align}  
where
\begin{align}
\label{h1h2}
H_1= {\tilde\pi_2+i\tilde\pi_1 \choose \sigma_0'-m-i\tilde\pi_3}, \quad
H_2= {\tilde\sigma_1-i\tilde\sigma_2 \choose - \sigma_3'+i\tilde\pi_0},
\end{align}  
and $m=\sqrt{m_0^2+m_3^2}\simeq m_t/\sqrt 2$. The diagonalization of the mixing between the neutral scalars $\sigma_0$ and $\sigma_3$ results in the mass eigenstates $\tilde\sigma_0$, $\tilde\sigma_3$ and is characterized by the mixing angle $\theta'\neq \theta$. Hence, in the weak doublets, these states are presented as a mixture with the angle $\alpha =\theta-\theta'$  that compensates a difference in the angles
\begin{align}
\label{h1h}
\sigma_0'= \tilde\sigma_0 \cos \alpha + \tilde\sigma_3 \sin \alpha , \\
\label{h2h}
\sigma_3'= \tilde\sigma_3 \cos \alpha - \tilde\sigma_0 \sin \alpha .
\end{align}  
It should be noted that in the new variables the nonzero vacuum expectation develops only the field $H_1$, $\langle H_1 \rangle = (0, -m)$. Such behaviour is typical of any model with two Higgs doublet states \cite{Branco12}, and is known as the Higgs basis $(H_1, H_2)$. 

It is appropriate to make a few comments. The first one concerns the form of the gap equations. There are many ways to rewrite them, but the following one is especially useful for calculations of the spectrum
\begin{align}
\label{extr1b}
&\bar g^2 \bar C_1= g_1- \frac{2g_2}{\cos 2\theta} +
  \frac{g_3}{\sin 2\theta}\, ,\\
\label{extr2b}
&\bar g^2 m^2\bar C_2= \frac{g_3}{\sin 2\theta}- \frac{g_2}{\cos
  2\theta}\, .
\end{align} 

Our second remark concerns the redefinition of Higgs fields. The fact is that the expression for the kinetic part of the free Higgs fields contained in (\ref{dl}) has a non-standard normalization factor
\begin{align}
\label{DH}
\mathcal L_H^{\mbox{\tiny kin}}&=\frac{C_2}{64\pi^2}  \, \mbox{tr}\left[\nabla_\mu \gamma^\mu
  (\sigma +i\gamma_5\pi )\right]^2 \nonumber \\
&=\frac{1}{2} \bar C_2\left( |D_\mu H_1|^2+|D_\mu H_2|^2 \right),  
\end{align} 
where the gauge covariant derivative is 
\begin{align}
D_\mu H_{1,2}=\left(\partial_\mu -i\, \frac{g}{2}T_iA_\mu^i-i\,\frac{g'}{2}B_\mu \right) H_{1,2}.
\end{align} 
To give (\ref{DH}) a canonical form, one rescales the fields $\tilde\pi_\alpha = \phi_\alpha / \sqrt{\bar C_2}$, and $\tilde\sigma_\alpha = \chi_\alpha / \sqrt{\bar C_2}$ in $H_{1,2}$. As a consequence, the two complex scalar doublets of the considered gauge theory are 
\begin{align}
\label{hh1}
&H_1= \frac{1}{\sqrt{\bar C_2}}{\phi_2+i\phi_1 \choose \chi_0'-\sqrt{\bar C_2}m-i\phi_3}, \\
\label{hh2}
&H_2=  \frac{1}{\sqrt{\bar C_2}} {\chi_1-i\chi_2 \choose - \chi_3'+i\phi_0}, 
\quad {\chi_0'  \choose \chi_3'}=R(\alpha) {\chi_0 \choose \chi_3},
\end{align}   
where the matrix $R(\alpha)$ is given by eq. (\ref{ot}). 

Hereinafter, along with variables $H_1$ and $H_2$, we use the set  
\begin{align}
\label{hh12}
\tilde H_1= {\phi_2+i\phi_1 \choose \chi_0'-i\phi_3}, \quad
\tilde H_2=  {\chi_1-i\chi_2 \choose - \chi_3'+i\phi_0}=\sqrt{\bar C_2}H_2,
\end{align}   
which is convenient, for example, when writing a number of specific vertices of an effective Lagrangian with explicitly written interaction constants.

To conclude this section, we write down the interaction part of the Higgs potential explicitly highlighting the terms of the third and fourth degree in the Higgs fields
\begin{align}
V_H^{\tiny int}&=\frac{2}{\bar C_2}\left[\frac{1}{4}(\tilde H_1^\dagger \tilde H_1+\tilde H_2^\dagger \tilde H_2)^2 + (\tilde H_1^\dagger \tilde H_1)(\tilde H_2^\dagger \tilde H_2) \right. \nonumber \\
&\left. - \left(\mbox{Im} ( \tilde H_1^\dagger \tilde H_2)\right)^2\right] -\frac{2m}{\sqrt{\bar C_2}} (\tilde H_1^\dagger \tilde H_1+\tilde H_2^\dagger \tilde H_2)\chi_0' \nonumber \\
&-\frac{4m_0}{\sqrt{\bar C_2}} (\cos \theta \chi_0' -\sin\theta \chi_3') \tilde H_2^\dagger \tilde H_2 
+ \frac{4m}{\sqrt{\bar C_2}} \phi_0 \, \mbox{Im} (\tilde  H_1^\dagger \tilde H_2)  \nonumber \\
&-\frac{4m_3}{\sqrt{\bar C_2}} (\cos \theta \chi_3' +\sin\theta \chi_0') \tilde H_1^\dagger \tilde H_1. 
\end{align}

\subsection{Nambu sum rule}

Let us discuss now the Higgs masses, which we derive from eq. (\ref{H}) (see also eq. (\ref{VHdiag}) and \cite{OsipovKh:20}). As one can see, the mass matrix eigenstates are $\chi_\alpha$ and $\phi_\alpha$ with the following squared values   
\begin{align}
\label{2lh}
&m_{\chi_0}^2=4m^2+\frac{2g_2}{\bar g^2 \bar C_2}\left(\frac{1}{\cos
  2\theta} -\frac{1}{\cos 2\theta'}  \right), \\
&m_{\chi_3}^2=4m^2+\frac{2g_2}{\bar g^2 \bar C_2}\left(\frac{1}{\cos
  2\theta} +\frac{1}{\cos 2\theta'} \right), \\
\label{sh}
&m^2_{\phi_0}=\frac{4g_2}{\bar g^2 \bar C_2\cos 2\theta }\, , \\
\label{shch}
&m^2_{\chi^\pm}=\frac{4g_3}{\bar g^2 \bar C_2\sin 2\theta }\, , \\
\label{gb}
&m^2_{\phi_i}=0. 
\end{align}
It follows that of the eight spinless states of the theory, three $\phi_i$ are massless Goldstone modes that are absorbed by gauge fields (Higgs mechanism). The remaining five, as one can easily verify
using (\ref{extr2b}), satisfy the sum rule
\begin{align}
\label{sr1a}
&m_{\chi_0}^2+m_{\chi_3}^2+m^2_{\phi_0}=\frac{8g_3}{\bar g^2 \bar C_2\sin
  2\theta }\, , \\
\label{sr2a}
&m^2_{\chi^+}+m^2_{\chi^-}=\frac{8g_3}{\bar g^2 \bar C_2\sin 2\theta }\, .
\end{align}
This result is somewhat different from the Nambu sum rule. Indeed, although the sum of the squared masses of the neutral modes and a similar sum for charged modes are equal to the same expression, its value is not $4m_t^2$, as required by the Nambu sum rule. In addition, instead of two Nambu partners, the first expression contains the contributions of three states, which also distinguishes this result from the standard pattern. In the following we clarify this behavior.

We address this matter, starting by writing down two relations, which are also a direct consequence of the mass formulas (\ref{2lh})-(\ref{gb}), namely
\begin{align}
\label{sr2}
&m_{\chi_0}^2+m_{\chi_3}^2=m^2_{\phi_0}+4(m_t^2 + m_b^2), \\
&m^2_{\chi^+}+m^2_{\chi^-}=2m^2_{\phi_0}+4(m_t^2 + m_b^2).
\end{align}
This shows that the non-zero mass of the $\phi_0$ meson is the only reason why the standard Nambu sum rule is violated.

As we have already noted, in the absence of the interaction with the coupling constant $g_2$, the theory possesses an additional $U(1)_A$ symmetry. It plays the role of the global symmetry of Peccei-Quinn \cite {Peccei77a,Peccei77b}, and prevents the appearance of the mass of the $\phi_0$ meson, which can be interpreted as an ``electroweak axion''. Indeed, one can verify that if $g_2 = 0$, the masses of the Higgs particles take the following values
\begin{align}
\label{lh2}
m_{\chi_0}=2m_b, \  m_{\chi_3}=2m_t, \  m_{\chi^\pm}=2m, \  m_{\phi_0}=0. 
\end{align}
Here we used (\ref{ta}) to obtain the ratio
\begin{align}
\label{lim}
\left. \frac{g_2}{\cos 2\theta '}\right|_{g_2=0}=2g_3, 
\end{align}
and the gap-equations (\ref{extr1c})-(\ref{extr2c}) to establish that at $g_2=0$ 
\begin{align}
\label{gap0}
2g_3=(m_t^2-m_b^2)\bar g^2 \bar C_2. 
\end{align}
The expressions (\ref {lh2}) are in a total agreement with the Nambu sum rule. Thus, one can conclude that the $U(1)_A$ breaking interaction $\propto g_2$ is responsible for the deviation from the canonical Nambu sum rule found in (\ref{sr1a})-(\ref{sr2a}).

One may wish to verify the quark content of the composite Higgs particles. For example, a neutral state with the mass $2m_b$ must be the $\bar bb$ bound state. Indeed, let us consider the field function that describes this state
\begin{align}
\label{wfh1}
\chi_0&\propto \tilde\sigma_0=\sigma_0 \cos\theta' +\sigma_3 \sin\theta'  \nonumber \\
&\propto \bar tt [(g_1+g_2+g_3)\cos\theta' +(g_1-g_2+g_3)\sin\theta'] \nonumber \\
&+\bar bb [(g_1+g_2-g_3)\cos\theta' -(g_1-g_2-g_3)\sin\theta' ] \nonumber \\
&\propto \bar bb.
\end{align}
Here the formula (\ref{qcsf}) has been used. On the last stage, we took into account that at $g_2 = 0$ and $m_0\neq m_3$ the angle $\theta '= - \pi /4 $, as it follows from (\ref{ta}). In the same manner, the quark content of the remaining states can be revealed.

Thus, in the model under consideration, the light composite Higgs boson is built mainly of $\bar bb$ quarks. Only due to the interaction $\propto g_2$ does an admixture of top quarks appear. That leads to an increase in its mass, which therefore occurs in the interval $2m_b<m_{\chi_0}<m_t$. 

Note, that in the limit of $m_{\chi_0}\ll\Lambda$ one can reliably use the renormalization group to improve the predictions for the low-energy Higgs masses obtained above by the resummation of the leading logarithmic corrections to arbitrary loop order. We postpone this analyses to the future. 

\subsection{Higgs-quark sector and Yukawa couplings} 

Consider the Yukawa part of the Lagrangian density (\ref{stL2}) which can be made into a somewhat more explicit formula for the Yukawa couplings  (see \ref{app5} for details)
\begin{align}
\label{hyc}
\mathcal L_Y=&\bar\psi \left(\sigma +i\gamma_5 \pi \right)\psi =\frac{1}{m}\left(m_b\bar\psi_L\tilde\Phi_1 b_R+m_t\bar\psi_L^a e_{ab} \tilde\Phi^*_{1b}t_R \right. \nonumber \\
&\left. +m_t\bar\psi_L\tilde\Phi_2 b_R-m_b\bar\psi_L^a e_{ab} \tilde\Phi^*_{2b}t_R  \right)  +h.c. 
\end{align}
The Higgs doublets used in this formula are given by eq. (\ref{H1andH2}). Here one should still take into account the effect of spontaneous electroweak symmetry breaking considering the non zero vacuum expectation values of scalar fields. One should also use the properly normalized Higgs fields (\ref{hh12}). As a result we come to the following expression  
\begin{align}
\label{hyc2}
\mathcal L_Y=&-\bar\psi M\psi + \lambda_b\bar\psi_L\tilde H_1b_R 
+\lambda_t\bar\psi_L^a e_{ab} \tilde H^*_{1b} t_R \nonumber \\
&+\lambda_t \bar\psi_L \tilde H_2 b_R-\lambda_b\bar\psi_L^a e_{ab} \tilde H^*_{2b}t_R +h.c. 
\end{align}
Here the completely antisymmetric unit tensor of second rank $e_{ab}$ has two nonzero components $e_{12}=-e_{21}=1$. This can also be represented by the matrix $i\tau_2$. The quark-mass-matrix $M=\mbox{diag} (m_t,m_b)$, and the Higgs-Yukawa couplings $\lambda_t$ and $\lambda_b$ are 
\begin{align}
\label{ltt}
\lambda_t=\frac{m_t}{m \sqrt{\bar C_2}}\equiv\frac{y_t}{\sqrt 2}, \quad
\lambda_b=\frac{m_b}{m \sqrt{\bar C_2}}\equiv\frac{y_b}{\sqrt 2}.
\end{align}
The compositeness boundary condition $\bar C_2=0$ at $\mu =\Lambda$ can be associated with the boundary conditions for the running coupling constants: $y_t(\mu ), y_b(\mu )\to\infty $ at the high-energy scale $\mu \to\Lambda$ when applying the renormalization group equations. Let us recall that the boundary condition states that if the Higgs doublet is a pure quark-antiquark bound state, then the corresponding Higgs-Yukawa coupling to the quarks must have a Landau pole at the composite scale $\Lambda$. The couplings $y_{t,b}(\mu )$ are only weakly sensitive to their initial values $y_{t,b}(\Lambda)$ because $\mu\ll \Lambda$ and, as a result, they have enough time to approach an infrared fixed point  \cite{Hill03}.

\subsection{Gauge bosons}
\label{Gauge}

Consider now the induced effective Lagrangian which describes the physics of gauge fields. Its kinetic part follows from eqs. (\ref{YM}) and (\ref{dl}) and is given by the density  
\begin{align}
\label{GB1}
\mathcal L_{\mbox{\tiny gauge}}^{\mbox{\tiny kin}}=\mathcal L_{\mbox{\tiny YM}} +\frac{\bar C_2}{96}\,\mbox{tr}\left(F_{\mu\nu}^2\right)+ \frac{\bar C_2}{256}\,\mbox{tr}\left([\gamma^\mu, \gamma^\nu ]F_{\mu\nu}\right)^2,
\end{align}
where the trace over color degrees of freedom is trivial: it gives the factor $N_c$ which is absorbed in $\bar C_2$. Thus, the symbol "tr" is understood here as a trace over $SU(2)$ tau matrices and Dirac gamma matrices. The last term originates from the $Y^2$ part in (\ref{dl}). We are not integrating over the gauge-boson fields and need specify no gauge fixing at this stage. The expression obtained can be simplified. The details are given in \ref{app6}. As a result we have     
\begin{align}
\label{GB2} 
\mathcal L_{\mbox{\tiny gauge}}^{\mbox{\tiny kin}}=\mathcal L_{\mbox{\tiny YM}} -\frac{\bar C_2}{48}\,\mbox{tr}\left(F_{\mu\nu}^2\right).
\end{align}  
Notice that the integration of short-distance components of quark fields induces a low-energy correction to the Yang-Mills Lagrangian density in the following form which deviates from the structure of the SM. Indeed, after evaluation of the trace 
\begin{align}
\mbox{tr}\left(F_{\mu\nu}^2\right)=2\mbox{tr}_f \left[g'^{\, 2} (Q^2+Y_L^2)B_{\mu\nu}^2+g^2 G_{\mu\nu}^2 \right], 
\end{align} 
where $G^{\mu\nu}=G^{\mu\nu}_i T_i$, and $Y_L=1/6$, we obtain 
\begin{align} 
\mathcal L_{\mbox{\tiny gauge}}^{\mbox{\tiny kin}}=\mathcal L_{\mbox{\tiny YM}}-\frac{g^2\bar C_2}{48}\left(\frac{11}{9}\tan ^2 \theta_W B_{\mu\nu}^2 + \vec G_{\mu\nu}^2\right).
\end{align}  
The whole expression can be written in terms of physical gauge fields as follows
\begin{align}
\mathcal L_{\mbox{\tiny gauge}}^{\mbox{\tiny kin}}=-\frac{1}{4}&\left[\left(\frac{1}{Z_B}+\frac{g^2\bar C_2}{12}\frac{11}{9}\tan ^2 \theta_W\right)B_{\mu\nu}^2 \right. \nonumber \\
&\left. + \left(\frac{1}{Z_A}+\frac{g^2\bar C_2}{12}\right)\vec G_{\mu\nu}^2\right].
\end{align}  

Now the free real parameters $Z_A$ and $Z_B$ can be fixed by requiring the standard form for kinetic terms of gauge-fields. This yields
\begin{align}
Z_A^{-1}=1-\frac{g^2\bar C_2}{12}, \quad Z_B^{-1}=1-\frac{g^2\bar C_2}{12}\frac{11}{9}\tan ^2 \theta_W.
\end{align}
Thus, by appropriate rescaling of the gauge fields, we were able to show that the form of the kinetic term does not change with the scale of $\mu$. 

This cannot be said about the masses of the gauge fields generated by the Higgs mechanism from the term (\ref{DH})
\begin{align}
\mathcal L_{\mbox{\tiny gauge}}^{\mbox{\tiny mass}}=\frac{1}{4}\,g^2m^2\bar C_2\left(W_\mu^+W_\mu^-+\frac{Z_\mu^2}{2\cos^2\theta_W}\right).
\end{align}
which depends on the scale through $\bar C_2$. The masses of the gauge fields are 
\begin{align}
\label{mgf}
m_W=\frac{1}{2}\, gm\, \bar C_2^{1/2} =m_Z\cos\theta_W.
\end{align}
This expression may be compared with the well-known SM result $m_W=gv/2$, where $v$ is the vacuum-expectation value of the standard Higgs field. Combining these formulas we find that
\begin{align}
\label{v}
v=m \sqrt{\bar C_2 } = 254.6\, \mbox{GeV}.
\end{align}
Finally, this helps to establish the ratio $\Lambda/\mu$:
\begin{align}
\label{nvl}
\frac{\Lambda}{\mu}= \exp\left[ \frac{(2\pi v)^2}{N_c(m_t^2+m_b^2)} \right]=2.345\times 10^{12}.
\end{align}
At scale $\mu = \Lambda_{EW}$, it gives $\Lambda \simeq 0.58\times 10^{15}\mbox{GeV}$.  
This suggest the possible role of GUT in the arising of four-quark interactions as a result of the GUT symmetry breaking. 


\section{Numerical estimates}
\label{NE}

The "minimal bosonization" model considered here has five free parameters: $g_1,  g_2, g_3, \Lambda, \mu$. We will fix them at the scale of the SM, i.e., at $\mu =\Lambda_{EW}=246\,\mbox{GeV}$. Then, in accord with (\ref{nvl}), we have $\Lambda = 0.58\times 10^{15}\mbox{GeV}$. We also assume that the couplings $g_1, g_2, g_3$ can be chosen in a way to obtain the phenomenologically consistent solutions of gap equations, i.e, one can fix these parameters to obtain the experimental values of quark masses, $m_t=173\pm 0.4\,\mbox{GeV}$, $m_b=4.18^{+0.04}_{-0.03}\,\mbox{GeV}$. Additionally, we require that the mass of the Higgs state is $m_{\chi_0}=125\,\mbox{GeV}$. Let us show now under which conditions our expectations can be fulfilled.

The spectrum of the Higgs states (\ref{2lh})-(\ref{shch}) depends only on the values of three independent parameters. These are the quark masses $m_t$ and $m_b$, and the ratio $g_3/g_2$, which we parametrise by introducing the dimensionless parameter $a$ 
\begin{align}
\label{a}
\frac{g_3}{g_2}=a\tan 2\theta . 
\end{align}
The other parameters in (\ref{2lh})-(\ref{shch}) can be eliminated with the help of the gap equation (\ref{extr2b}). Let us also remind that the angle $\theta$ and the mass parameter $m$ in Eqs.(\ref{2lh})-(\ref{shch}) are functions of the quark masses $m_t$ and $m_b$    
\begin{align}
\tan 2\theta = \frac{m_t^2-m_b^2}{2m_tm_b}, \quad m^2=\frac{1}{2}\left(m_t^2+m_b^2\right).
\end{align}
That gives $\theta=43.6^\circ$. 

The angle $\theta'$ can be expressed only in terms of the masses of quarks and the parameter $a$. Moreover, from Eq.(\ref{ta})
\begin{align}
\label{thetaprime}
\tan 2\theta' =(3-2a)\tan 2\theta,
\end{align}
we conclude that the angle $\theta'<0$ if $a>3/2$.   

Now we can represent the mass formulae (\ref{2lh})-(\ref{shch}) in the form
\begin{align}
\label{h1}
&m_{\chi_0}^2=\frac{2m^2}{a-1} \left( 2a-1 - \Delta \right), \\
\label{h2}
&m_{\chi_3}^2=\frac{2m^2}{a-1} \left( 2a-1 +\Delta \right), \\
\label{hn}
&m^2_{\phi_0}=\frac{4m^2}{a-1}\, , \\
\label{hpm}
&m^2_{\chi^\pm}=\frac{4m^2 a}{a-1}\, , 
\end{align}
where
\begin{align}
\label{D}
  \Delta =\sqrt{\cos^22\theta +(3-2a)^2\sin^2 2\theta}.
\end{align}
If we fix the parameter $a$ according to the known mass value of the standard Higgs state: $m_{\chi_0} = 125 \, \mbox{GeV}\to a = 4.84$, then the following estimates are obtained from these formulas: $m_{\chi_3}=346\,\mbox{GeV}, m_{\chi^\pm}=275\,\mbox{GeV}, m_{\phi_0}=125\,\mbox{GeV}$, and for the angle $\theta'$ we find from eq.(\ref{thetaprime}) the value $\theta' =-44.8^\circ$.

Let us derive the values of four-Fermi couplings. From eq. (\ref{a}) we have $g_3/g_2 =10^2$, so $g_3\gg g_2$. It means, that one can resort to the gap-equations (\ref{extr1c})-(\ref{extr2c}) to estimate the values of $g_1$ and $g_3$. These equations show that the coupling constants of the model must be extremely fine-tuned when $\mu^2\ll \Lambda^2$. Indeed, we have
\begin{eqnarray}
g_1&=&g_c+\mathcal O(\frac{\mu^2}{\Lambda^2}), \quad g_c=\frac{4\pi^2}{N_c\Lambda^2}= 3.9\times 10^{-29}\,\mbox{GeV}^{-2},
\nonumber \\
g_3&=&g_c \frac{m_t^2-m_b^2}{2\Lambda^2}\ln\frac{\Lambda^2}{\mu^2} +\mathcal O(\frac{\mu^2}{\Lambda^2}). 
\end{eqnarray}
This is, indeed, the usual fine-tuning or the gauge-hierarchy problem of the SM, which is isolated in the gap equation sector of the NJL approach. Our estimates imply extreme proximity of $g_1$ to the critical value $g_c$, and $g_3$ to the value $ 2.5 g_c \times 10^{-24}\simeq 1.0\times 10^{-52}\,\mbox{GeV}^{-2}$.  Thus, the couplings must be fine-tuned to within 
$$ \frac{g_1}{g_c}:\frac{g_2}{g_c}:\frac{g_3}{g_c} \sim 1:10^{-26}:10^{-24}$$
of the critical value $g_c$.

In conclusion, we note that despite a satisfactory description of both the quark masses and the ground Higgs state, the predictions obtained for the neutral, $\phi_0$, and charged, $\chi^\pm$, Higgs states are most likely experimentally disfavoured. A more refined calculation of the mass spectrum based on the renormalization group approach can shift these values. Work in this direction is in progress.

\section{Conclusions}

In this paper we concentrated on the vacuum structure of the model proposed by Miransky, Tanabashi and Yamawaki in \cite{Mir89}. To make our calculations as transparent as possible we restricted to the "minimal bosonization" procedure, which does not generate new coupling constants. We also used the Schwinger-DeWitt method in a form that fully meets the problem of deriving the induced effective Lagrangian, the induced parameters of which must have an explicit dependence upon two scales $\Lambda$ and $\mu$, vanishing when $\mu\to\Lambda$ (compositeness condition).  

The minimal bosonization procedure leads to the specific quark-meson part of the Lagrangian which possesses the chiral $U(2)_L\times U(2)_R$ symmetry. Our reasoning was to avoid  calculations of the fermion one loop diagrams directly in the Nambu - Goldstone phase, where fermions have very different masses $m_t\gg m_b$, which are not easy to handle consistently. It turned out to be more efficient to start from the massless fermion loop in the symmetric phase. In this way, we managed to obtain a rather simple effective Lagrangian, to extract its extremum conditions (gap equations), to specify the ground state of the theory, and, finally, to analyze the main consequences of the approach for the spectrum of collective modes, which are the composite Higgs states.    

This approach has a series of interesting consequences. Firstly it leads to a phenomenological value for the mass of the SM composite Higgs. This result is interesting because, as we have already noted, top-condensation models usually yield significantly overestimated values. We have shown that the standard Higgs is not a pure $\bar tt$ bound state, but has an essential part associated with the light bottom quarks. The underlying mechanism is obvious from eq. (\ref{wfh1}). The estimates showed that the angle $\theta'$ is close to $-\pi/4$, and that therefore the $\bar bb$ component is dominant in the Higgs $\chi_0$ field.

Secondly, we investigated the question of whether the Nambu sum rules are satisfied in the model with two Higgs doublets. The presence of the fifth neutral boson $\phi_0$ violates the standard picture of Nambu partners. The initially massless "electroweak axion" acquires its mass as a result of $U(1)_A$ symmetry breaking ('t Hooft interaction). This violates the Nambu sum rules. Since  an interaction with a much weaker coupling constant $g_2$ (as compared to $g_1$)  is necessary for the generation of the mass of this state, we conclude that the standard form of the Nambu sum rules is valid only in the leading order in $1/N_c$. It is surprising that such a violation leads to a degeneracy in mass for the main Higgs $\chi_0$ and $\phi_0$, $m_{\chi_0}\simeq m_{\phi_0}$. This degeneracy is of a random nature and we believe it will be removed after taking into account the quantum corrections.

Taking into account the electroweak and strong corrections is still necessary for a complete picture of the emerging spectrum of states. Such work is under way, and we hope that it will not greatly affect the result presented here.

The key problem with the model considered is that the new dynamics lies at a very high energy scale $\Lambda \sim 10^{15}\,\mbox{GeV}$. This value corresponds to the GUT scale $10^{15}\,\mbox{GeV}$ giving some credit to the scenarios where new interactions are generated by GUT physics. On the other hand, this shows that the model is extremely fine-tuned. This is the known hierarchy problem of the SM. The top condensation models cannot clarify this question. However, we saw that the fine-tuning problem  is isolated in the gap equations. Once we tune couplings $g_1, g_2, g_3$ to admit the desirable solution no further quadratic divergences in other amplitudes need be canceled.
     
We hope that the results obtained here contribute to the further development of the model, since they are obtained on the basis of fairly simple approximations which, on one hand, have been quite successful and, on the other, can be developed in several straightforward directions. 

\section*{Acknowledgments}
A. A. Osipov would like to thank C.T. Hill for his interest in this study and valuable correspondence. This Research is supported by CFisUC and Funda\ca o para a Ci\^encia e Tecnologia through the project UID/FIS/04564/ 2016, and grant CERN/FIS-COM/0035/2019. F. Palanca acknowledges the M.Sc. grant by Instituto Cam\~oes and INAGBE, J. Moreira a research grant by project POCI-01-0145-FEDER-029912, and  M. Sampaio a research grant  from CNPq. We  would like acknowledge networking support by the COST Action CA16201.

\appendix

\section{Linearization of four-Fermi interactions}\label{app1}

To prove the dynamical equivalence of the theories based on the Lagrangian densities (\ref{stL1}) and (\ref{sembos}) one should integrate over auxiliary fields in the functional integral. Since the auxiliary fields appear quadratically in the functional integral they can be integrated out exactly. For that one should solve the eight Euler-Lagrange equations for auxiliary fields $\sigma_\alpha$, $\pi_\alpha$ regarding the quark bilinear combinations $\bar\psi \tau_\alpha \psi$ and  $\bar\psi i\gamma_5\tau_\alpha\psi$.  So, what one has to make sure to prove dynamical equivalence of the two theories is that (\ref{stL1}) and (\ref{sembos}) coincide on the extremal trajectories, given by the Euler-Lagrange equations
\begin{equation}
\frac{\partial\mathcal L'}{\partial\sigma_\alpha}=0, \quad \frac{\partial\mathcal L'}{\partial\pi_\alpha}=0.
\end{equation}    
The solution of this system of linear equations is straightforward
\begin{align}  
\label{qcsf} 
&2\sigma_0=(g_1 +g_2)\bar\psi\tau_0\psi + g_3 \bar\psi\tau_3\psi , \nonumber \\
&2\sigma_1=(g_1 -g_2)\bar\psi\tau_1\psi - g_3 \bar\psi i\gamma_5\tau_2\psi , \nonumber \\
&2\sigma_2=(g_1 -g_2)\bar\psi\tau_2\psi +g_3\bar\psi i\gamma_5\tau_1\psi , \nonumber \\
&2\sigma_3=(g_1 -g_2)\bar\psi\tau_3\psi + g_3 \bar\psi\tau_0\psi , \nonumber \\
&2\pi_0=(g_1 -g_2)\bar\psi i\gamma_5\tau_0\psi +g_3\bar\psi  i\gamma_5\tau_3\psi , \nonumber \\
&2\pi_1=(g_1 +g_2)\bar\psi i\gamma_5\tau_1\psi +g_3\bar\psi \tau_2\psi , \nonumber \\
&2\pi_2=(g_1 +g_2)\bar\psi i\gamma_5\tau_2\psi -g_3\bar\psi \tau_1\psi , \nonumber \\
&2\pi_3=(g_1 +g_2)\bar\psi i\gamma_5\tau_3\psi +g_3\bar\psi  i\gamma_5\tau_0\psi . 
\end{align}
One can see that quark-antiquark bound states $\sigma_0, \sigma_3$ are described by the linear combination of scalars, the $\pi_0, \pi_3$ are the mixture of two pseudoscalars. The other four bound states are the mixture of the scalar and pseudoscalar quark-antiquark bilinears.  

Inserting these solutions in (\ref{sembos}) yields  
\begin{align}
&\bar\psi (\sigma +i\gamma_5 \pi )\psi +\mathcal L_{\pi, \sigma} =\frac{1}{2} \bar\psi (\sigma +i\gamma_5 \pi )\psi \nonumber \\
&=\frac{g_1}{4} \left[(\bar\psi\tau_\alpha\psi)^2+(\bar\psi i\gamma_5\tau_\alpha\psi)^2\right] \nonumber \\
&+\frac{g_2}{4}\left[(\bar\psi\tau_0\psi)^2-(\bar\psi\tau_i\psi)^2 - (\bar\psi i\gamma_5\tau_0\psi)^2 +(\bar\psi i\gamma_5\tau_i\psi)^2   \right] \nonumber \\
& +\frac{g_3}{2}\left[(\bar\psi\tau_0\psi)(\bar\psi\tau_3\psi)- (\bar\psi\tau_1\psi) (\bar\psi i\gamma_5\tau_2\psi) \right. \nonumber \\
&\left. \quad\ \  + (\bar\psi\tau_2\psi) (\bar\psi i\gamma_5\tau_1\psi) +(\bar\psi i\gamma_5\tau_0\psi) (\bar\psi i\gamma_5\tau_3\psi)\right]. 
\end{align}

What we need now is to demonstrate that this expression matches exactly with $\mathcal L_{4\psi}$, given by eq. (\ref{MTY}). For this purpose let us use the left and right chiral components of the Dirac field $\psi$ 
\begin{align}
&\bar \psi\tau_\alpha\psi =\tau_\alpha^{ab}\left(\bar\psi^a_L\psi_R^b+\bar\psi^a_R\psi^b_L\right)
\equiv s_\alpha , \\
&\bar \psi i\gamma_5\tau_\alpha\psi = i \tau_\alpha^{ab}\left(\bar\psi^a_L\psi_R^b-\bar\psi^a_R\psi^b_L\right)\equiv p_\alpha ,
\end{align}     
and define the complex $2\times 2$ matrix 
\begin{align}
&\Sigma_{ab}=\frac{1}{2}\left(s_\alpha +i p_\alpha \right)\tau_\alpha^{ab}=2\bar\psi_R^b\psi_L^a, \\
&\Sigma_{ab}^\dagger =\frac{1}{2}\left(s_\alpha -i p_\alpha \right)\tau_\alpha^{ab}=2\bar\psi_L^b\psi_R^a.
\end{align}

Then one can easily see that the first term of eq. (\ref{MTY}) can be written as  
\begin{align}
\label{1term}
(\bar\psi_L^a\psi_R^b)(\bar\psi_R^b\psi_L^a)&=\frac{1}{4}\Sigma^\dagger_{ba}\Sigma_{ab}=\frac{1}{4}\mbox{tr}\left(\Sigma^\dagger\Sigma\right) =\frac{1}{8}(s_\alpha^2+p_\alpha^2) \nonumber \\
&=\frac{1}{8} \left[(\bar\psi\tau_\alpha\psi)^2+(\bar\psi i\gamma_5\tau_\alpha\psi)^2\right]
\end{align}

To convert the second term in  (\ref{MTY}) we need a well-known formula
\begin{align}
(i\tau_2)^{ab}(i\tau_2)^{cd}=\epsilon_{ab}\epsilon_{cd}, 
\end{align}
which gives 
\begin{align}
&(\bar\psi_L^a\psi_R^c) (i\tau_2)^{ab}(i\tau_2)^{cd} (\bar\psi_L^b\psi_R^d)=\frac{1}{4}\Sigma^\dagger_{ca}\Sigma_{db}^\dagger \epsilon_{ab}\epsilon_{cd} =\frac{1}{2}\det \Sigma^\dagger \nonumber \\
&=\frac{1}{8}\left[s_0^2 -s_i^2 -p_0^2+p_i^2-2i(s_0p_0-s_ip_i)\right].
\end{align}
It follows then that  
\begin{align}
\label{2term}
&(\bar\psi_L^a\psi_R^c) (i\tau_2)^{ab}(i\tau_2)^{cd} (\bar\psi_L^b\psi_R^d) + h.c.     \\
&=\frac{1}{2}\left(\det \Sigma^\dagger + \det\Sigma\right)=\frac{1}{4}\left[s_0^2 -s_i^2 -p_0^2+p_i^2\right] \nonumber \\
&=\frac{1}{4}\left[(\bar\psi\tau_0\psi)^2-(\bar\psi\tau_i\psi)^2 - (\bar\psi i\gamma_5\tau_0\psi)^2 +(\bar\psi i\gamma_5\tau_i\psi)^2   \right]. \nonumber 
\end{align}

For the third term in (\ref{MTY}) we have 
\begin{align}
\label{3term}
&(\bar\psi_L^a\psi_R^b) \tau_3^{bc}(\bar\psi_R^c\psi_L^a)=\frac{1}{4}\Sigma^\dagger_{ba} \tau_3^{bc} \Sigma_{ac}=\frac{1}{4}\mbox{tr}\left(\Sigma^\dagger \Sigma \tau_3 \right) \nonumber \\
&=\frac{1}{4}\left(s_0s_3  -s_1p_2+s_2p_1+p_0p_3\right) \nonumber \\
&=\frac{1}{4}\left[(\bar\psi\tau_0\psi)(\bar\psi\tau_3\psi)- (\bar\psi\tau_1\psi) (\bar\psi i\gamma_5\tau_2\psi) \right. \nonumber \\
&\left. \quad\ \  + (\bar\psi\tau_2\psi) (\bar\psi i\gamma_5\tau_1\psi) +(\bar\psi i\gamma_5\tau_0\psi) (\bar\psi i\gamma_5\tau_3\psi)\right]. 
\end{align}
Combining the formulas (\ref{1term}), (\ref{2term}) and (\ref{3term}) one can see the equivalence of Lagrangian densities (\ref{stL1}) and (\ref{sembos}).

\section{Infinitesimal transformations of fields}\label{app2}

We start from the infinitesimal $U(2)_V\times U(2)_A$ transformations of quark fields
\begin{equation}
\label{tr1}
\delta\psi =i(\alpha +\gamma_5\beta )\psi, \quad \delta\bar \psi =i\bar\psi(-\alpha +\gamma_5\beta ),
\end{equation}
where $\alpha =\frac{1}{2}\alpha_a \tau_a$, $\beta =\frac{1}{2}\beta_a \tau_a$, and $\alpha_a, \beta_a$ are the eight $(a =0,1,2,3)$ infinitesimal parameters of the global transformations of the vector $U(2)_V$ and axial-vector $U(2)_A$ groups. 

The invariance of the Yukawa term $\bar\psi (\sigma +i\gamma_5\pi)\psi$ under transformations (\ref{tr1}) means the validity of the following chiral transformation laws for bosonic fields
\begin{equation}
\delta\sigma =i [\alpha , \sigma ]+\{\beta ,\pi\},\quad \delta\pi =i[\alpha ,\pi ]-\{\beta ,\sigma\}.
\end{equation}    
Since mesonic fields are real functions, the transformations do not depend on $\alpha_0$. Thus, we deal with the $SU(2)_V\times U(2)_A$ group. Commutators and anti-commutators can be calculated by using the basic properties of Pauli-matrices. This yields
\begin{equation}
[\tau_i, \tau_j ]=2i e_{ijk} \tau_k, \quad \{\tau_a, \tau_b \}=2h_{abc}\tau_c,
\end{equation} 
where $i,j,k =1,2,3$ and the only non zero components of the totally symmetric coefficients $h_{abc}$ are $h_{000}=1, h_{0ij}=\delta_{ij}$. Then we have 
\begin{eqnarray}
&&\delta\sigma_a = \frac{1}{2} \mbox{tr} \left(\tau_a \delta\sigma\right)=h_{abc}\beta_b\pi_c -\delta_{ak}e_{kij}\alpha_i\sigma_j \nonumber \\
&&\delta\pi_a = \frac{1}{2} \mbox{tr} \left(\tau_a \delta\pi\right)=-h_{abc}\beta_b\sigma_c -\delta_{ak}e_{kij}\alpha_i\pi_j
\end{eqnarray}
or, in the components
\begin{eqnarray}
\label{components}
&&\delta\sigma_0=\beta_a\pi_a, \quad \delta\sigma_i=-e_{ijk}\alpha_j\sigma_k +\beta_0\pi_i+\beta_i\pi_0, \nonumber \\
&&\delta\pi_0=- \beta_a\sigma_a, \quad \delta\pi_i=-e_{ijk}\alpha_j\pi_k -\beta_0\sigma_i-\beta_i\sigma_0 ,
\end{eqnarray}

The formulas (\ref{components}) allow us to establish how the group $SU(2)_V\times U(2)_A$ acts on the following complex combinations of fields
\begin{align}
\label{uu2tr}
\delta (\sigma_0 -i\pi_3)&=i\left[ \alpha_1\pi_2-\alpha_2\pi_1+ \beta_0\sigma_3+\beta_3\sigma_0 -i\beta_a\pi_a\right], \nonumber \\
\delta (\pi_2 +i\pi_1)&= i\left[ -(\alpha_1-i\alpha_2)i\pi_3-(\beta_1-i\beta_2)\sigma_0 \right. \nonumber \\&\left. \ \ \ \ \ \ +\alpha_3(\pi_2+i\pi_1) -\beta_0(\sigma_1-i\sigma_2) \right], \nonumber \\
\delta (i\pi_0-\sigma_3)&= \alpha_1\sigma_2-\alpha_2\sigma_1 -\beta_0\pi_3-\beta_3\pi_0-i\beta_a\sigma_a , \nonumber\\
\delta (\sigma_1-i\sigma_2)&=i\left[\alpha_3(\sigma_1-i\sigma_2) - (\alpha_1-i\alpha_2)\sigma_3 \right. \nonumber \\
&\left. -\beta_0 (\pi_2+i\pi_1) -\pi_0(\beta_2+i\beta_1)\right].
\end{align}

From these formulas, in particular, one can find how these combinations are transformed under the action of the group $SU(2)_L\times U(1)_A$. To project the infinitesimal action of the chiral group $SU(2)_V\times U(2)_A$ on the $SU(2)_L\times U(1)_A$ group one should properly identify the set of relevant infinitesimal parameters. In the case of $SU(2)_L$ one should chose the left transformations, i.e., to put the parameters of right-hand transformations to zero. Requiring that $\alpha_i +\beta_i=0$, we arrive to the set of four parameters $\omega_i/2\equiv\alpha_i=-\beta_i, \ \beta_0$ which parametrize the infinitesimal action of the group. From (\ref{uu2tr}) one finds the action of the group on the bosonic fields
\begin{align}
\label{su2L}
\delta (\sigma_0 -i\pi_3)&=\frac{i}{2}\left[ (\omega_1+i\omega_2)(\pi_2+i\pi_1) -\omega_3(\sigma_0 -i\pi_3)\right] \nonumber \\
&- i\beta_0 (i\pi_0-\sigma_3), \nonumber \\
\delta (\pi_2 +i\pi_1)&=\frac{i}{2}\left[ (\omega_1-i\omega_2)(\sigma_0-i\pi_3) +\omega_3(\pi_2 +i\pi_1)\right] \nonumber \\
&- i\beta_0 (\sigma_1-i\sigma_2), \nonumber \\
\delta (i\pi_0-\sigma_3)&= \frac{i}{2}\left[ (\omega_1+i\omega_2)(\sigma_1-i\sigma_2) -\omega_3(i\pi_0-\sigma_3)\right] \nonumber \\
&- i\beta_0 (\sigma_0-i\pi_3), \nonumber \\
\delta (\sigma_1-i\sigma_2)&=\frac{i}{2}\left[ (\omega_1-i\omega_2)(i\pi_0-\sigma_3) +\omega_3(\sigma_1-i\sigma_2)\right] \nonumber \\
&- i\beta_0 (\pi_2+i\pi_1), 
\end{align}
These formulas show that $\Phi_1$ and $\Phi_2$ in (\ref{phidoublets}) behave like fundamental $SU(2)_L$-doublets, and that $U(1)_A$ transformations exchange them
\begin{align}
\delta \Phi_{1,2}=i\omega_i\,\frac{\tau_i}{2}\, \Phi_{1,2}-i\beta_0\Phi_{2,1}. 
\end{align} 
We may consider the $U(1)_A$ symmetry in a different basis
\begin{align}  
\Phi_1=\frac{1}{\sqrt 2}(\Phi_1' +\Phi_2'), \quad \Phi_2=\frac{1}{\sqrt 2}(\Phi_2' -\Phi_1'),
\end{align}
in which the axial group acts without mixing of fields, but changing sign
\begin{align}
\delta \Phi_{1,2}'=i\omega_i\,\frac{\tau_i}{2}\, \Phi_{1,2}'\pm i\beta_0\Phi_{1,2}'. 
\end{align}
This $U(1)$ symmetry (suitably extended to the quark sector) was first introduced by Peccei and Quinn \cite{Peccei77a,Peccei77b} in connection with the strong-CP problem.

\section{Quark content of the Higgs fields}\label{app3}

Let us clarify the quark content of the scalar auxiliary variables $\sigma_\alpha$ and $\pi_\alpha$. This helps to understand the origin of the quark masses and the reasoning for the specific grouping of these fields into the electroweak doublets. Our starting point are eqs. (\ref{qcsf}), which can be investigated for the existence of solutions homogeneous in space-time. In particular, it gives
\begin{align}
&2\langle\sigma_0\rangle =(g_1+g_2+g_3)\langle\bar tt\rangle +(g_1+g_2-g_3)\langle\bar bb\rangle ,\nonumber \\
&2\langle\sigma_3\rangle =(g_1-g_2+g_3)\langle\bar tt\rangle -(g_1-g_2-g_3)\langle\bar bb\rangle .
\end{align}
Expressed in terms of top and bottom quark masses they take the form
\begin{align}
&m_t =-(g_1+g_3)\langle\bar tt\rangle -g_2 \langle\bar bb\rangle ,\nonumber \\
&m_b =-g_2 \langle\bar tt\rangle -(g_1-g_3)\langle\bar bb\rangle .
\end{align}
It follows then that even in the absence of the bottom quark condensate, $\langle\bar bb\rangle$, the presence of top condensate, $\langle\bar tt\rangle$, may generate the mass of both quarks, provided that $U(1)_A$ symmetry is violated, $g_2\neq 0$.    

From eqs. (\ref{qcsf}) one can also obtain that  
\begin{align}
&\sigma_0-i\pi_3 = (g_1+g_2+g_3)\bar t_Lt_R +(g_1+g_2-g_3)\bar b_Rb_L, \nonumber \\
&\pi_2 +i\pi_1= (g_1+g_2-g_3)\bar b_Rt_L-(g_1+g_2+g_3)\bar b_Lt_R, \nonumber \\
&i\pi_0-\sigma_3=(g_1-g_2-g_3)\bar b_Rb_L -(g_1-g_2+g_3)\bar t_Lt_R, \nonumber \\
&\sigma_1 -i\sigma_2= (g_1-g_2-g_3)\bar b_Rt_L+(g_1-g_2+g_3)\bar b_Lt_R. 
\end{align}
These relations show that if one neglects the mixing generated by the interaction with the coupling $g_3$, in other words if one puts $g_3=0$, two electroweak doublets in eq. (\ref{phidoublets}) would have the following quark content 
\begin{align}
\label{quarkcont}
&\Phi_1= (g_1+g_2) { \bar b_Rt_L-\bar b_Lt_R  \choose \bar b_Rb_L+\bar t_Lt_R},  \\
&\Phi_2= (g_1-g_2) { \bar b_Rt_L+\bar b_Lt_R \choose \bar b_Rb_L-\bar t_Lt_R}.
\end{align}  
The upper states here have the positive charge $Q=1$ and $T_3=1/2$, the lower ones are neutral, $Q=0$, and $T_3=-1/2$. Therefore both doublets can be characterized by the U(1) hypercharge $Y_L=Q-T_3=1/2$.  

\section{Diagonalization of the Higgs states}\label{app4}

The arbitrary quadratic form
\begin{align}
\label{qf}
 \Omega^x_y (a, b, c/2 )= (x,y)  
 \left( \begin{array}{cc}
                             a  & c/2 \\
                             c/2 & b
                             \end{array}
                      \right)
   {x \choose y}
\end{align}
can be diagonalized by the orthogonal transformation $R(\theta)$ to the new variables $(\tilde x, \tilde y)$
\begin{equation}
\label{ot}
   {x \choose y}=\left( \begin{array}{cc}
                             \cos\theta  &\sin\theta \\
                             -\sin\theta &\cos\theta
                             \end{array}
                      \right)
   {\tilde x \choose \tilde y} =R(\theta ) {\tilde x \choose \tilde y}.
\end{equation}
The condition 
\begin{align}
\tan 2\theta =\frac{c}{b-a}
\end{align}
nullifies off-diagonal terms. The diagonal ones are
\begin{align}
& \Omega^x_y (a, b, c/2 )= \Omega^{\tilde x}_{\tilde y} (a_{11}, a_{22},0), \nonumber \\
&a_{11}= \frac{1}{2}\left[a+b +(a-b)\sqrt{1+\tan^2 2\theta}  \right], \nonumber \\
&a_{22}= \frac{1}{2}\left[a+b -(a-b)\sqrt{1+\tan^2 2\theta}  \right].
\end{align}
With these notations the quadratic part of the potential (\ref{VH}) after the shifts $\sigma_0\to\sigma_0 -m_0 $, and $\sigma_3\to\sigma_3 -m_3 $ can be written   
\begin{align}
V^{(2)}_H=\Omega^{\pi_0}_{\pi_3} +\Omega^{\sigma_3}_{\sigma_0} +\Omega^{\pi_2}_{\sigma_1}+\Omega^{\sigma_2}_{\pi_1}, 
\end{align}
where arguments of the quadratic forms are given correspondingly
\begin{align}
& a_{11}^{\pi_0,\pi_3}=\frac{1}{\bar g^2} (g_1+g_2)+m^2 \bar C_2- \bar C_1, \nonumber \\
& a_{22}^{\pi_0,\pi_3}=\frac{1}{\bar g^2}(g_1-g_2) +m^2 \bar C_2- \bar C_1, \nonumber \\
& a_{12}^{\pi_0,\pi_3}=2m_0m_3 \bar C_2 -\frac{g_3}{\bar g^2}.
\end{align}
\begin{align}
& a_{11}^{\sigma_3,\sigma_0}=\frac{1}{\bar g^2} (g_1+g_2)+3m^2 \bar C_2- \bar C_1, \nonumber \\
& a_{22}^{\sigma_3,\sigma_0}=\frac{1}{\bar g^2}(g_1-g_2) +3m^2 \bar C_2- \bar C_1, \nonumber \\
& a_{12}^{\sigma_3,\sigma_0}=6m_0m_3 \bar C_2 -\frac{g_3}{\bar g^2}.
\end{align}
\begin{align}
& a_{11}^{\pi_2,\sigma_1}=\frac{1}{\bar g^2} (g_1-g_2)+(m_0^2+3m_3^2)\bar C_2-\bar C_1,\nonumber \\
& a_{22}^{\pi_2,\sigma_1}=\frac{1}{\bar g^2} (g_1+g_2)+(3m_0^2+m_3^2)\bar C_2-\bar C_1,\nonumber \\
& a_{12}^{\pi_2,\sigma_1}=\frac{g_3}{\bar g^2}.
\end{align}
\begin{align}
& a_{11}^{\sigma_2,\pi_1}=\frac{1}{\bar g^2} (g_1+g_2)+(3m_0^2+m_3^2)\bar C_2-\bar C_1,\nonumber \\
& a_{22}^{\sigma_2,\pi_1}=\frac{1}{\bar g^2} (g_1-g_2)+(m_0^2+3m_3^2)\bar C_2-\bar C_1,\nonumber \\
& a_{12}^{\sigma_2,\pi_1}=-\frac{g_3}{\bar g^2},
\end{align}
and we recall that $m^2=m_0^2+m_3^2$. 

The diagonalization of these forms show that 
\begin{align}
\theta_{\pi_0\pi_3}=\theta_{\pi_2\sigma_1}=\theta_{\sigma_2\pi_1}\equiv \theta, 
\end{align}
where 
\begin{align}
\tan 2\theta =\frac{m_t^2-m_b^2}{2m_tm_b}, \quad \mbox{or} \quad \tan\theta =\frac{m_3}{m_0}.
\end{align}
To get these results we have used gap equations (\ref{extrem1})-(\ref{extrem2}). 

In the case of $\Omega^{\sigma_3}_{\sigma_0}$ we obtain the different angle
\begin{align}
\theta_{\sigma_3\sigma_0}\equiv \theta ', 
\end{align}
with the relation 
\begin{align}
\tan 2\theta' =3\frac{m_t^2-m_b^2}{2m_tm_b} -2\frac{g_3}{g_2}=3\tan 2\theta  -2\frac{g_3}{g_2}. 
\end{align}

The diagonalized quadratic part of the potential $V_H$ is given by  
\begin{align}
\label{VHdiag}
V_H^{(2)}&=\frac{g_2}{\bar g^2} \frac{m_t^2+m_b^2}{m_tm_b}\tilde\pi_0^2 + \frac{2g_3}{\bar g^2} \frac{m_t^2+m_b^2}{m_t^2-m_b^2}\left(\tilde\sigma_1^2+\tilde\sigma_2^2\right)  \\
&+\left[(m_t^2+m_b^2)\bar C_2 + \frac{g_2}{\bar g^2}\left(\frac{1}{\cos 2\theta}-\frac{1}{\cos 2\theta'}\right) \right] \tilde\sigma_0^2 \nonumber \\
&+\left[(m_t^2+m_b^2)\bar C_2 + \frac{g_2}{\bar g^2}\left(\frac{1}{\cos 2\theta}+\frac{1}{\cos 2\theta'}\right) \right] \tilde\sigma_3^2. \nonumber
\end{align}

It is easy to establish that states with a definite mass (described by the variables which diagonalize $V^{(2)}_H$) are associated with the initial doublets (\ref{phidoublets}) by the orthogonal transformation 
\begin{equation}
\label{otph}
   {\Phi_1 \choose \Phi_2}= \left( \begin{array}{cc}
                             \cos\theta  &\sin\theta \\
                             -\sin\theta &\cos\theta
                             \end{array}
                      \right)
   {\tilde\Phi_1 \choose \tilde\Phi_2},
\end{equation}
where the rotated states are collected in the two doublets
\begin{equation}
\label{H1andH2}
\tilde\Phi_1= {\tilde \pi_2+i\tilde\pi_1  \choose \sigma_0'-i\tilde \pi_3}, \quad 
\tilde\Phi_2= {\tilde \sigma_1-i \tilde \sigma_2  \choose -\sigma_3'+i\tilde \pi_0}.
\end{equation}
Here $\sigma_0'$ and $\sigma_3'$ are linear combinations of the neutral scalar physical states $\tilde\sigma_0, \tilde\sigma_3$ 
\begin{equation}
\label{neutral}
   {\sigma_3' \choose \sigma_0'}\equiv \left( \begin{array}{cc}
                             \cos\alpha  & -\sin\alpha \\
                             \sin\alpha &\cos\alpha
                             \end{array}
                      \right)
   {\tilde\sigma_3 \choose \tilde\sigma_0}.
\end{equation}
The angle $\alpha =\theta -\theta'$. 

The obtained formulas represent a solution to the problem of diagonalizing the quadratic part of the potential $V_H$, i.e. it is assumed that the vacuum state is correctly determined. However, If we are interested in the problem of minimizing the potential $V_H$ and want to examine the ground state of the theory, the formulas  (\ref{otph}) should be modified by substitutions $\sigma_0\to\sigma_0-m_0$, and $\sigma_3\to\sigma_3-m_3$ in accord with eq. (\ref{h1h2}). This replacement does not change $\tilde\Phi_2 = H_2$, because
\begin{align}
{\sigma_3' \choose \sigma_0'}&\to {\sigma_3' \choose \sigma_0'} - R(-\alpha) R(-\theta') {m_3 \choose m_0} \nonumber \\
&={\sigma_3' \choose \sigma_0'} - R(-\theta) {m_3 \choose m_0}={\sigma_3' \choose \sigma_0'} - {0 \choose m},
\end{align}
where, on the last stage, we used the relations
\begin{align}
\cos\theta = \frac{m_0}{m}, \quad \sin\theta = \frac{m_3}{m}.
\end{align}
However, $\tilde\Phi_1$ is changed to $H_1$. As a consequence, the lower component of $H_1$ has a (real and negative) vacuum expectation value $-m$, while $H_2$ has a null expectation value.

\section{The Yukawa part of the Higgs Lagrangian}\label{app5}

Let us consider the Yukawa part of the Lagrangian density (\ref{stL2}) 
\begin{align}
&\mathcal L_Y=\bar\psi (\sigma + i\gamma_5 \pi) \psi =\bar\psi_L (\sigma + i\pi)\psi_R + \bar\psi_R (\sigma - i\pi)\psi_L \nonumber \\
&=\bar t_L t_R (\sigma_0+\sigma_3 +i\pi_0+i\pi_3) 
+\bar t_L b_R (\sigma_1-i\sigma_2 +i\pi_1+\pi_2) \nonumber \\
&+\bar b_L b_R (\sigma_0-\sigma_3 +i\pi_0-i\pi_3) 
+\bar b_L t_R (\sigma_1+i\sigma_2 +i\pi_1-\pi_2) \nonumber \\
&+ h.c. 
\end{align}
This can be written in terms of physical fields
\begin{align}
\mathcal L_Y &=\frac{1}{m}\left\{
\bar t_L t_R [m_t (\sigma_0' +i\tilde\pi_3)+m_b(\sigma_3'+i\tilde\pi_0 )] \right. \nonumber \\
&+\bar t_L b_R [m_t (\tilde\sigma_1 -i\tilde\sigma_2)+m_b(\tilde\pi_2+i\tilde\pi_1)] \nonumber \\
&+\bar b_L b_R [m_b (\sigma_0' -i\tilde\pi_3)+m_t(-\sigma_3'+i\tilde\pi_0 )] \nonumber \\
&\left. +\bar b_L t_R [m_b (\tilde\sigma_1 +i\tilde\sigma_2)-m_t(\tilde\pi_2-i\tilde\pi_1)] + h.c. \right\}. 
\end{align}

Noting that 
\begin{align}
&\bar\psi_L \tilde\Phi_1 b_R= \bar t_L b_R (\tilde\pi_2+i\tilde\pi_1)+\bar b_L b_R (\sigma_0' -i\tilde\pi_3), \nonumber \\
&\bar\psi_L \tilde\Phi_2 b_R= \bar t_L b_R (\tilde\sigma_1-i\tilde\sigma_2)-\bar b_L b_R (\sigma_3' -i\tilde\pi_0), \nonumber \\
&\bar\psi^a_L e_{ab} \tilde\Phi_1^{*b} t_R=\bar t_L t_R(\sigma_0' +i\tilde\pi_3)-\bar b_L t_R (\tilde\pi_2-i\tilde\pi_1) \nonumber \\
&\bar\psi^a_L e_{ab} \tilde\Phi_2^{*b} t_R=-\bar t_L t_R(\sigma_3' +i\tilde\pi_0)-\bar b_L t_R (\tilde\sigma_1+i\tilde\sigma_2), 
\end{align}  
where $\tilde\Phi_1$ and $\tilde\Phi_2$ are given by (\ref{H1andH2}), $e_{ab}$ is totally antisymmetric tensor with $e_{12}=1$, we find 
\begin{align}
\label{Yukawa}
\mathcal L_Y&=\frac{1}{m}\left( m_b \bar\psi_L \tilde\Phi_1 b_R + m_t \bar\psi_L \tilde\Phi_2 b_R \right.  \nonumber \\
& \left. +m_t \bar\psi^a_L e_{ab} \tilde\Phi_1^{*b} t_R  - m_b \bar\psi^a_L e_{ab} \tilde\Phi_2^{*b} t_R + h.c. \right).
\end{align}

\section{An useful formula}\label{app6}

Here we present mathematical details related with the step made from eq.(\ref{GB1}) to eq.(\ref{GB2}). Namely, we prove the equality 
\begin{align}
\label{formula}
\mbox{tr}\left([\gamma^\mu, \gamma^\nu ]F_{\mu\nu}\right)^2 =-8 \,\mbox{tr}\left(F_{\mu\nu}\right)^2 + \partial_\mu V^\mu.
\end{align}
The four-divergence of a four-vector $V^\mu$ does not contribute to the variation of the action and hence does not affect the dynamical characteristics of the system. We recall that "tr" means the calculation of two traces tr =tr$_D$ tr$_f$ taken with respect to the Dirac matrices, tr$_D$, and $SU(2)$ flavour matrices, tr$_f$.  

The spin-1 fields strength tensor $F_{\mu\nu}$ is defined by eqs. (\ref{F})-(\ref{Gm}). Its gauge-field content is
\begin{align}
F^{\mu\nu}=P_R F_R^{\mu\nu} + P_L F_L^{\mu\nu},
\end{align}  
where 
\begin{align}
F_{R}^{\mu\nu} = g' Q B^{\mu\nu}, \quad  F_{L}^{\mu\nu} = g G^{\mu\nu} +g' Y_L B^{\mu\nu},
\end{align}
and 
\begin{align}
G^{\mu\nu}=\partial^\mu A^\nu -\partial^\nu A^\mu -ig [A^\mu, A^\nu].
\end{align}
Using these formulas we find 
\begin{align}
&\mbox{tr} \left( 
[\gamma^\mu, \gamma^\nu ] F_{\mu\nu} \right)^2 \nonumber \\
&=\mbox{tr}\left\{
[\gamma_\mu, \gamma_\nu ][\gamma_\rho, \gamma_\sigma ] \left(P_R F_{R}^{\mu\nu} F_{R}^{\rho\sigma} + P_L F_{L}^{\mu\nu} F_{L}^{\rho\sigma} \right) \right\} \nonumber \\
&= \frac{1}{2}\mbox{tr}\left\{
[\gamma_\mu, \gamma_\nu ][\gamma_\rho, \gamma_\sigma ] \left(F_{R}^{\mu\nu} F_{R}^{\rho\sigma} + F_{L}^{\mu\nu} F_{L}^{\rho\sigma} \right) \right\} \nonumber \\
&+ \frac{1}{2}\mbox{tr}\left\{
[\gamma_\mu, \gamma_\nu ][\gamma_\rho, \gamma_\sigma ] \gamma_5 \left(F_{R}^{\mu\nu} F_{R}^{\rho\sigma} - F_{L}^{\mu\nu} F_{L}^{\rho\sigma} \right) \right\} \nonumber \\
&=-16\,\mbox{tr}_f\left(F_{R}^{\mu\nu} F_{R}^{\mu\nu} +F_{L}^{\mu\nu} F_{L}^{\mu\nu} \right) \nonumber \\
&+8i e_{\mu\nu\rho\sigma} \mbox{tr}_f \left(F_{R}^{\mu\nu} F_{R}^{\rho\sigma} - F_{L}^{\mu\nu} F_{L}^{\rho\sigma} \right).
\end{align}

It is straightforward to see now that the first term here gives the first term on the right hand side of eq.(\ref{formula}). Indeed,
\begin{align}
\mbox{tr}\left(F_{\mu\nu}\right)^2&=\mbox{tr} \left(P_R F_{R}^{\mu\nu} F_{R}^{\mu\nu} + P_L F_{L}^{\mu\nu} F_{L}^{\mu\nu} \right) \nonumber \\
&=2 \,\mbox{tr}_f \left(F_{R}^{\mu\nu} F_{R}^{\mu\nu} + F_{L}^{\mu\nu} F_{L}^{\mu\nu} \right).
\end{align}

Let us consider now the second term. Our goal is to show that it has a form of a total derivative. The calculations show that
\begin{align}
e_{\mu\nu\rho\sigma} \mbox{tr}_f F_{R}^{\mu\nu} F_{R}^{\rho\sigma} =  \mbox{tr}_f \partial^\mu \left( 4 g'^{\,2} Q^2 e_{\mu\nu\rho\sigma} B^\nu\partial^\rho B^\sigma  \right), 
\end{align}
\begin{align}
&e_{\mu\nu\rho\sigma} \mbox{tr}_f F_{L}^{\mu\nu} F_{L}^{\rho\sigma} = e_{\mu\nu\rho\sigma} \mbox{tr}_f  \left[ g^2 G^{\mu\nu}  G^{\rho\sigma} +  g'^{\,2} Y_L^2 B^{\mu\nu}B^{\rho\sigma}\right] \nonumber \\
&=4 e_{\mu\nu\rho\sigma} \mbox{tr}_f  \partial^\mu \left(g^2  A^\nu\partial^\rho A^\sigma- \frac{2i}{3} g^3 A^\nu A^\rho A^\sigma \right. \nonumber \\
&\left. \qquad \qquad \qquad \  +g'^{\,2} Y_L^2 B^\nu\partial^\rho B^\sigma \right) 
\end{align}
Therefore, we find
\begin{align}
V_\mu =32 i e_{\mu\nu\rho\sigma} \mbox{tr}_f & \left[ g'^{\,2}(Q^2-Y_L^2) B^\nu\partial^\rho B^\sigma -g^2A^\nu\partial^\rho A^\sigma  \right. \nonumber \\
&\left. + \frac{2i}{3} g^3 A^\nu A^\rho A^\sigma \right]. 
\end{align}
Thus, the validity of the formula (\ref{formula}) is established.


\end{document}